\documentclass[a4paper,11pt]{article}
\usepackage{pos}
\usepackage{amsmath}
\usepackage{graphicx}
\usepackage{appendix}
\usepackage{xcolor}
\usepackage{url}

\title{Reconciling the kinematical constraint with the JIMWLK evolution equation: correlation functions non-local in rapidity}

\author*[a]{P. Korcyl}
\author[a]{L. Motyka}
\author[a]{T. Stebel}

\affiliation[a]{Institute of Theoretical Physics, Jagiellonian University, \\ ul. \L ojasiewicza 11, 30-348 Krak\'ow, Poland }

\emailAdd{piotr.korcyl@uj.edu.pl}
\emailAdd{leszek.motyka@uj.edu.pl}
\emailAdd{tomasz.stebel@uj.edu.pl}

\abstract{    In the high-energy limit of DIS experiments the effective degrees of freedom of QCD are Wilson line operators. Their evolution in the rapidity variable is predicted by the set of Balitsky-JIMWLK evolution equations. We analyze a new class of two-point correlation functions of Wilson line operators where the Wilson lines are taken at different values of the rapidity variable. Such correlation functions can appear in the discussion of the kinematical constraint. We find that in the Langevin formulation of the JIMWLK equation, such correlation functions are affected by an infrared divergence. We discuss a possible regularization of this divergence and its consequences for the implementation of the kinematical constraint for the JIMWLK equation.}

\FullConference{31st International Workshop on Deep Inelastic Scattering (DIS2024)\\
 8–12 April 2024\\
Grenoble, France\\}

\begin{document}

\maketitle

\section{Introduction}
\label{sec:intro}

The description of a number of observables accessible in the high-energy limit of Deep Inelastic Scattering experiments requires knowledge of correlation functions of multiple Wilson line operators. Unless some approximation is used to bring them down to the well-known two-point dipole gluon amplitude, this fact entails the use of a more general, finite-$N_c$ evolution equation, i.e. the set of Balitsky-JIMWLK evolution equations. An example of a recent calculation where this is the case is provided by the estimation of the di-jet cross-section in the nearly back-to-back limit \cite{Caucal:2023nci,Caucal:2023fsf} where the Weizs\"acker-Williams distribution is needed. In this contribution, we describe the numerical setup suitable for the solution of the B-JIMWLK equation and discuss the progress in implementing the kinematical constraint into that framework. The latter, first introduced in Refs.\cite{KC1_Andersson:1995ju, KC2_Kwiecinski:1996td,KC3_Salam:1998tj}, was implemented into the momentum-space BK equation in Ref.\cite{KC4_Golec-Biernat:2001dqn} and later in the position-space BK equation in Refs.\cite{KC5_Motyka:2009gi,KC6_Beuf:2014uia,KC7_Iancu:2015vea}. Ref.\cite{Hatta:2016ujq} describes a construction that implements the kinematical constraint into the Langevin framework for the B-JIMWLK equation \cite{Langevin1_Weigert_2002,Langevin2_Blaizot_2003,Langevin3_Rummukainen_2004}. Basic building blocks of this proposal are correlation functions of Wilson lines at different rapidities. In this contribution, we continue the studies of such functions obtained numerically by solving the B-JIMWLK equation. In Ref.\cite{Korcyl:2024xph} we have defined two variants of correlators where one Wilson line is at a different rapidity than the other and derived their BK-type evolution equations. It allowed us to identify the infrared divergence present in these equations. In the present contribution, we propose a possible regularization of the divergence and discuss its effect on the dynamics of the full kinematically constraint B-JIMWLK evolution equation.

\section{Correlation functions of two Wilson lines with different rapidities}

\begin{figure}
\begin{center}
    \includegraphics[width=0.75\textwidth]{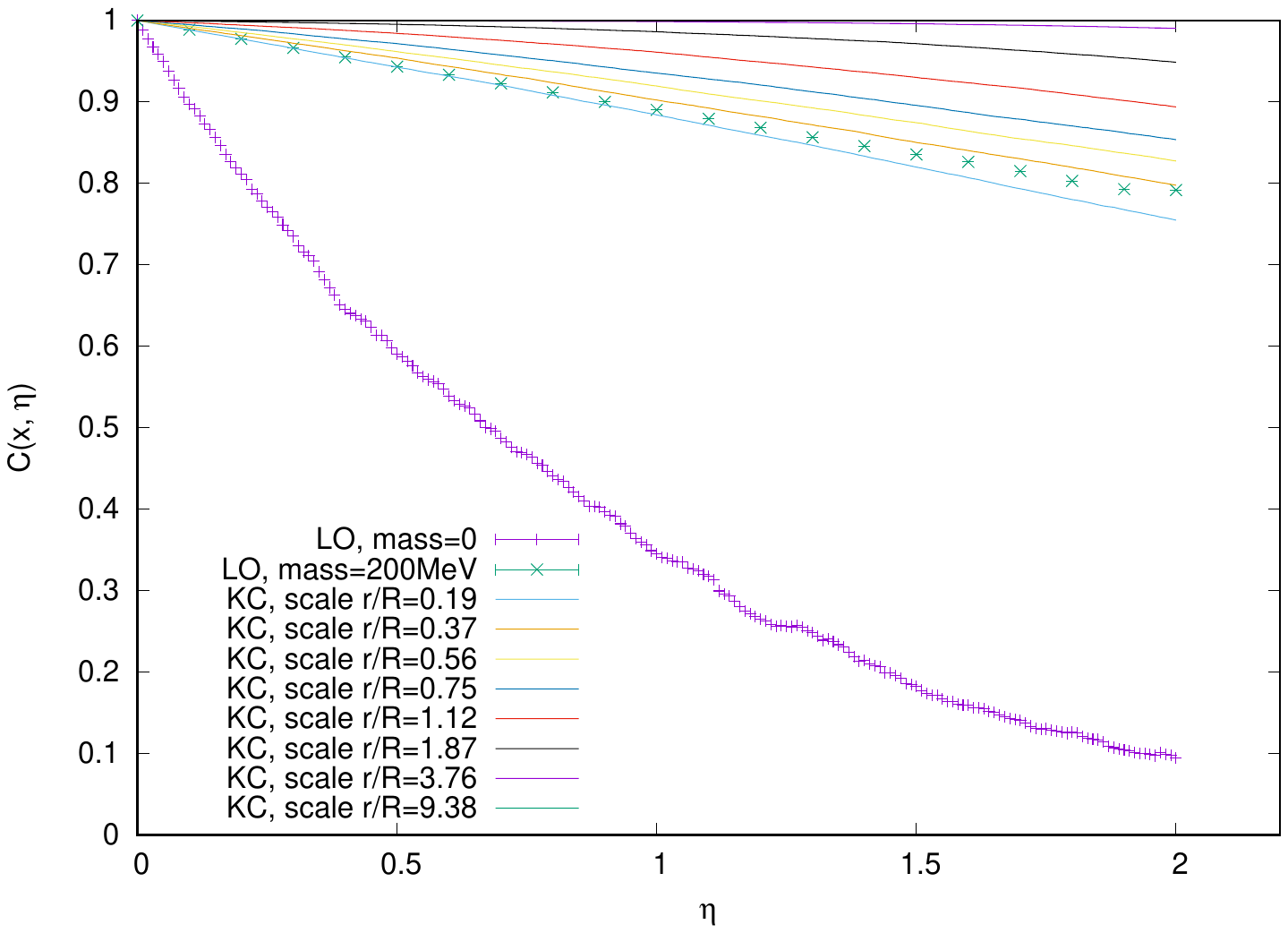}
    \caption{The correlation functions $C(\mathbf{x},\eta)$ evaluated in different setups. Both violet pluses and green crosses were obtained with the LO JIMWLK evolution equation, but we used the value of the parameter $m$ corresponding to $200$ MeV for the latter. The solid lines were obtained with the KC JIMWLK evolution equation at the same value of $m$ and different scales $r/R=r/R_{\textrm{init}}$, where $R_{\textrm{init}}$ is the saturation radius of the initial condition which sets the scale to the problem.  \label{fig. 1}}
\end{center}
\end{figure}

The construction of Ref.\cite{Hatta:2016ujq} introduces two-point functions of Wilson lines which are defined with different rapidities. We have provided explicit definitions of these function in our previous proceedings Ref.\cite{Korcyl:2024xph}. The single-point correlation function reads
\begin{equation}
\label{eq. rapidity correlator}
    C(\mathbf{x}, \eta) = \frac{1}{N_c} \langle \textrm{tr} U^{\dagger}(\mathbf{x}, 0) U(\mathbf{x}, \eta) \rangle_{\textrm{CGC}}.
\end{equation}
and the more general correlation two-point function $W(r,\eta)$ is
\begin{equation}
\label{eq. W correlator}
    W(\mathbf{r},\eta) = \frac{1}{N_c} \frac{1}{V} \sum_{\mathbf{x}} \langle \textrm{tr} U^{\dagger}(\mathbf{x}, 0) U(\mathbf{x}+\mathbf{r}, \eta) \rangle_{\textrm{CGC}}.
\end{equation}
The large-$N_c$ evolution equations can be derived using the leading-order B-JIMWLK equation for the evolution of a single Wilson line in $\eta$,
\begin{align}
    \frac{\partial W(\mathbf{x}-\mathbf{y},\eta)}{\partial \eta} 
    &= \frac{\bar{\alpha}_s}{2 \pi} \int_{\mathbf{z}} \mathcal{K}_{xz}\Big(  S(\mathbf{x}-\mathbf{z},\eta) W(\mathbf{z}-\mathbf{y},\eta) - W(\mathbf{x}-\mathbf{y},\eta) \Big), \label{eq. W}\\
    \frac{\partial C(\eta)}{\partial \eta} 
      &= \frac{\bar{\alpha}_s}{2 \pi} \int_{\mathbf{z}} \mathcal{K}_{xz} \Big(
        S(\mathbf{x}-\mathbf{z},\eta) W(\mathbf{z}-\mathbf{x},\eta)  -C(\mathbf{x}, \eta) \Big), \label{eq. C}\\
    \frac{\partial S(\mathbf{x}-\mathbf{y},\eta))}{\partial \eta} &=  \frac{\bar{\alpha}_s}{2 \pi}    \int_{\mathbf{z}} \mathcal{M}_{xyz} \Big(
      S(\mathbf{x}-\mathbf{z},\eta) S(\mathbf{z}-\mathbf{y},\eta) - S(\mathbf{x}-\mathbf{y},\eta)\Big), \label{eq. normal bk eq}
\end{align}
where 
\begin{equation}
    K^i_{xz} = \frac{(\mathbf{x} - \mathbf{z})^i}{(\mathbf{x} - \mathbf{z})^2}, \textrm{ and } \mathcal{K}_{xz} =  K^i_{xz} K^i_{xz} = \frac{1}{|\mathbf{x}-\mathbf{z}|^2},
\end{equation}
and 
\begin{equation}
    \mathcal{M}_{xyz} \equiv  K^i_{xz} K^i_{xz} + K^i_{yz} K^i_{yz}  
-  2K^i_{xz}  K^i_{yz}
\end{equation}
and $S(\mathbf{x},\eta)$ is the usual gluon dipole amplitude with both Wilson lines taken at rapidity $\eta$. In Ref.\cite{Korcyl:2024xph} we identified an infrared divergence in these equations. It can be already seen in the first step of evolution for the $C$ correlation function since the initial slope is given by
\begin{equation}
    \frac{\partial C(\mathbf{x},\eta)}{\partial \eta}\Big|_{\eta=0} = \frac{\bar{\alpha}_s}{2 \pi} \int_{\mathbf{z}} \mathcal{K}_{xz} \Big(
        \exp(-|\mathbf{x}-\mathbf{z}|^2) -1 \Big),
\end{equation}
where we assumed a Gaussian initial condition for simplicity. The integral over the constant piece in the bracket is not converging on an infinite two-dimensional plane. In the numerical setup, the infinite plane is replaced by a torus of finite volume in order to maintain translational symmetry. Thus, the divergence is regulated. However, the larger the volume the faster $C$ and $W$ decay. Hence, the numerical results strongly depend on the size of the simulated torus which is undesirable. Simulation data from the full LO JIMWLK evolution equation confirm these conclusions. As shown in Fig.\ref{fig. 1} by the violet pluses the function $C(\eta)$ decays very quickly to zero. The simulations were performed in a volume of the linear extent of $96 R_{\textrm{init}}$, where $R_{\textrm{init}}$ is the saturation radius of the initial condition which sets the scale to the problem. A much better approach would be to regularize the large distance behavior of the kernels $\mathcal{K}$ and $\mathcal{M}$  in such a way that the results become independent of the volume of the torus.

\section{Adding a regularization parameter}

We follow Ref.\cite{Gardi:2006rp} and regularize the large distance behaviour of the kernels $\mathcal{K}$ and $\mathcal{M}$ with an additional parameter $m$,
\begin{equation}
    K^i_{xz} = \frac{(\mathbf{x} - \mathbf{z})^i}{|\mathbf{x} - \mathbf{z}|^2} \rightarrow \frac{(\mathbf{x} - \mathbf{z})^i}{|\mathbf{x} - \mathbf{z}|^2}  e^{-m |\mathbf{x} - \mathbf{z}| } \textrm{ then }     \mathcal{K}_{xz} =  K^i_{xz} K^i_{xz} \rightarrow \frac{1}{|\mathbf{x}-\mathbf{z}|^2} e^{-2m |\mathbf{x} - \mathbf{z}| } \label{eq. impr K}
\end{equation}
and
\begin{equation}
    \mathcal{M}_{xyz} \rightarrow 
    \frac{\big( (y-z)e^{-m |\mathbf{x} - \mathbf{z}| }-(x+z)e^{-m |\mathbf{y} - \mathbf{z}| } \big)^2}{(x-z)^2(y-z)^2}.  \label{eq. impr M}
\end{equation}
The motivation behind this modification is that the non-vanishing large distance behavior of the leading order kernel is not physical. We expect that the improved kernels Eqs.\eqref{eq. impr K} and \eqref{eq. impr M} will regulate the mentioned infrared divergence and the correlation functions $C$ and $W$ will not decorrelate so fast with $\eta$. In consequence, the assumptions of the implementation of the kinematical constraint from Ref.\cite{Hatta:2016ujq} should be better satisfied.

\section{Numerical results}

In Fig.\ref{fig. 1} we show the dependence of $C(\eta)$ on rapidity obtained using different approaches. The violet pluses correspond to the LO B-JIMWLK evolution equation as discussed in Ref.\cite{Korcyl:2024xph} and show the fast decorrelation. On the contrary, the green crosses were obtained using the same LO B-JIMWLK evolution equation but with the kernels given by Eqs.\eqref{eq. impr K} and \eqref{eq. impr M} and the value of the $m$ parameter around $200$ MeV. We observe that the decorrelation is indeed much slower and $C(\eta) \approx 1$. 

Additionally, we also plot the function $C(\eta)$ evaluated using the kinematically constrained B-JIMWLK equation using the prescription of Ref.\cite{Hatta:2016ujq} for different scales with solid lines. The data show that for the scales from a wide range between $0.2 R_{\textrm{init}}$ and $10 R_{\textrm{init}}$ the corresponding $C$ functions decay slowly with $\eta$. However, for the smallest of the scale $0.2 R_{\textrm{init}}$ $C$ decorrelates \emph{faster} than the LO B-JIMWLK results with the same value of the $m$ parameter. As a consequence, the gluon dipole amplitude calculated with the kinematically constrained B-JIMWLK equation evolves faster than the LO one, which is against our expectations.

\section{Conclusions}

We studied the dynamics of various correlation functions appearing in the B-JIMWLK evolution equation with the kinematical constraint. We concentrated on the correlation functions in which the Wilson lines are at different rapidities. We have identified and regulated an infrared divergence affecting their evolution by modifying the large distance behavior of the JIMWLK kernel. Although, as expected the decorrelation of these correlation functions becomes much slower, in some cases, it is still faster than that for the LO B-JIMWLK equation. We conclude that the construction of the kinematical constraint for the B-JIMWLK equation needs further studies. 

\section{Acknowledgements}
We gratefully acknowledge Polish high-performance computing infrastructure PLGrid (HPC Center: ACK Cyfronet AGH) for providing computer facilities and support within computational grants no. PLG/2022/015321 and no. PLG/2023/016656. T.S. kindly acknowledges the support of the Polish National Science Center (NCN) grant No. 2021/43/D/ST2/03375. P.K. acknowledges support from of the Polish National Science Center (NCN) grant No. 2022/46/E/ST2/00346.

\bibliographystyle{apsrev}
\bibliography{biblio.bib}

\end{document}